\documentstyle[psfig,12pt]{article}
\def\href#1#2{#2}

\def\beq{\begin{equation}}
\def\eeq{\end{equation}}

\begin{document}
\baselineskip=14.5pt \pagestyle{plain} \setcounter{page}{1}
\renewcommand{\thefootnote}{\fnsymbol{footnote}}
\newcommand{\da}{\dot{a}}
\newcommand{\db}{\dot{b}}
\newcommand{\dn}{\dot{n}}
\newcommand{\dda}{\ddot{a}}
\newcommand{\ddb}{\ddot{b}}
\newcommand{\ddn}{\ddot{n}}
\newcommand{\pa}{a^{\prime}}
\newcommand{\pb}{b^{\prime}}
\newcommand{\pn}{n^{\prime}}
\newcommand{\ppa}{a^{\prime \prime}}
\newcommand{\ppb}{b^{\prime \prime}}
\newcommand{\ppn}{n^{\prime \prime}}
\newcommand{\fda}{\frac{\da}{a}}
\newcommand{\fdb}{\frac{\db}{b}}
\newcommand{\fdn}{\frac{\dn}{n}}
\newcommand{\fdda}{\frac{\dda}{a}}
\newcommand{\fddb}{\frac{\ddb}{b}}
\newcommand{\fddn}{\frac{\ddn}{n}}
\newcommand{\fpa}{\frac{\pa}{a}}
\newcommand{\fpb}{\frac{\pb}{b}}
\newcommand{\fpn}{\frac{\pn}{n}}
\newcommand{\fppa}{\frac{\ppa}{a}}
\newcommand{\fppb}{\frac{\ppb}{b}}
\newcommand{\fppn}{\frac{\ppn}{n}}
\newcommand{\A}{A}
\newcommand{\B}{B}
\newcommand{\mmu}{\mu}
\newcommand{\mnu}{\nu}
\newcommand{\ii}{i}
\newcommand{\jj}{j}
\newcommand{\jl}{[}
\newcommand{\jr}{]}
\newcommand{\ml}{\sharp}
\newcommand{\mr}{\sharp}
\newcommand{\zol}{z}
\newcommand{\sqg}{\sqrt{-g}}
\newcommand{\diag}{\mbox{diag}}

\begin{flushright}
UW/PT-05-08\\
SLAC-PUB-11090\\
SU-ITP-05/12\\
IPM/P-2005/020\\
 {\tt hep-th/0504056}
\end{flushright}

\vskip 2cm

\begin{center}
{\Large \bf Hologravity
}
\vskip 1cm

{\bf Mohsen Alishahiha$^1$, Andreas Karch$^2$, and Eva
Silverstein$^3$}
\vskip 0.5cm

{\it $^1$ Institute for Studies in Theoretical Physics and
Mathematics, \\ P.O. Box 19395-5531, Tehran, Iran}\\
{\tt E-mail:alishah@ipm.ir} \\
\medskip
{\it $^2$ Department of Physics, University of Washington, \\
Seattle, WA 98195}
{\tt E-mail:karch@phys.washington.edu} \\
\medskip
{\it $^3$ SLAC and Department of Physics, Stanford
University, \\ Stanford, CA 94305/94309}
{\tt E-mail:evas@stanford.edu} \\

\end{center}

\vskip1cm

\begin{center}
{\bf Abstract}
\end{center}
\medskip

The dS/dS correspondence provides a holographic description of quantum gravity in $d$ dimensional de Sitter
space near the horizon of a causal region in a well defined approximation scheme; it is equivalent to the low
energy limit of conformal field theory on de Sitter space in $d-1$ dimensions coupled to $d-1$ dimensional
gravity.  In this work, we extend the duality to higher energy scales by performing calculations of various
basic physical quantities sensitive to the UV region of the geometry near the center of the causal patch.  In
the regime of energies below the $d$ dimensional Planck scale but above the curvature scale of the geometry,
these calculations encode the physics of the $d-1$ dimensional matter plus gravity system above the crossover
scale where gravitational effects become strong. They exhibit phenomena familiar from studies of two dimensional
gravity coupled to conformal field theory, including the cancellation 
of the total Weyl anomaly in
$d-1$ dimensions.  We also outline how the correspondence can be used to address the issue of observables in de
Sitter space, and generalize the correspondence to other space times, such as black holes, inflationary
universes, and landscape bubble decays.  In the cases with changing cosmological constant, we obtain a dual
description in terms of renormalization group flow.


\newpage

\section{Introduction}

In \cite{dsds} we established a holographic duality between gravity on $d$-dimensional de Sitter space (dS$_d$)
and a CFT on dS$_{d-1}$ below the scale of the inverse curvature radius $1/L$,  coupled to $d-1$ dimensional
gravity.  In this paper, we extend the duality above the scale $1/L$, by a set of controlled computations in the
bulk $d$-dimensional theory. These computations sum up effects of the residual gravity coupled to matter in
$d-1$ dimensions. We also describe qualitatively a formulation of the system based on further holographic
reduction, and a dual description of backgrounds with changing cosmological constant (as in inflation and
landscape decays) in terms of renormalization group flow.

The basic observation in \cite{dsds} was that when written in a dS$_{d-1}$ slicing, dS$_d$ has the form of a
Randall-Sundrum system \cite{rs2} with a smooth, built in analogue of the UV brane cutting off the theory in the
UV.  The metric for the dS$_d$ static patch can be written
\beq \label{slicing} ds^2_{dS_d}={1\over \cosh^2({z\over L})}(ds^2_{dS_{d-1}}+dz^2)     \eeq

Close to the horizon the warped geometry of dS$_d$ is isomorphic to the dS$_{d-1}$ slicing of AdS$_d$ (for which
the $1/\cosh^2(z/L)$ in the warped metric (\ref{slicing}) is replaced by $1/\sinh^2(z/L)$). For latter one has a
holographic duality in terms of a conformal field theory on dS$_{d-1}$. Hence given the UV/IR correspondence in
the AdS/CFT dictionary, we can conclude that at low energies, the dS causal patch is equivalent to a CFT on
dS$_{d-1}$.

In AdS$_{d-1}$, the warp factor diverges towards the UV region of
the geometry (far away from the horizon) and $d-1$ dimensional
gravity decouples.  In our case, as in Randall-Sundrum, the warp
factor is bounded in the solution and one finds a dynamical $d-1$
dimensional graviton. In the Randall-Sundrum construction, one
truncates the warp factor at a finite value of the radial
coordinate by including a brane source (or a compactification
manifold) with extra degrees of freedom.  In the dS case, the
additional brane source is unnecessary; a smooth UV brane at which
the warp factor turns around is built in to the geometry
\cite{autolocalize}.

In this paper, we make basic computations determining how the lower dimensional theory behaves above the scale
$1/L$. Our $d$ dimensional gravity calculations provide a controlled formulation of the system which translates
in the $d-1$ dimensional language (via a simple conformal map relating our system to one with AdS asymptotics)
to an effective description in which the total (matter plus gravity) central charge is zero and the operator
dimensions are reduced from their field theoretic values. These results are reminiscent of features familiar in
two dimensional gravity--namely gravitational dressing of operators and absence of a conformal anomaly including
the effects of gravity.  While these qualitative conclusions are to be expected, the values of the resulting
operator dimensions remain somewhat mysterious, as we will discuss further after explaining the computations. In
general, our results sum up the combined effects of $d-1$ dimensional gravity, matter, and their interactions
between the scale $1/L$ and the bulk Planck scale $M_d$.

Some of our results here also apply to the Randall-Sundrum system, providing a bulk computation of some of the
effects of induced gravity in that more general context. However, the dS case appears especially simple--in
particular the absence of an external brane source gives rise to the possibility that in the case of the dS/dS
correspondence, unlike in RS, no new degrees of freedom have to be introduced in the dual conformal field theory
at the scale $1/L$. The interplay of the two conformal field theories with the localized graviton may completely
fix the dual and naturally give rise to the physics of the crossover scale $1/L$, though further computations
would be required to test this.

After repeated application of the same duality down to 1+1 or 0+1
dimensional gravity, the holographic dual is a well defined, UV
complete theory. Basically it boils down to a standard quantum
mechanical system with a Hamiltonian constraint. As a result, the
observables of the full de Sitter space are similar to those in
toy models of quantum gravity obtained by truncation to
minisuperspace and reduction to one dimensional quantum mechanics,
albeit with a much richer matter sector.

Finally, we show that decays in cosmological constant correspond to a time dependent RG flow in our setup,
similarly to \cite{andy} but here in a system with a Lorentzian-signature CFT sector.  In particular, in our
setup the time direction is common in both the $d$ and $d-1$ descriptions, so it is particularly straightforward
to incorporate time dependent physics.  The bulk gravitational physics corresponding to the region near the
horizon translates into physics of the matter sector, in this case yielding a relation between inflation and a
time dependent RG flow.

In the following section we will review in more detail the scales that are involved in the dS/dS correspondence.
We will also examine repeated applications to 1+1 or 0+1 dimension, allowing one to define dS quantum gravity
via the observables of a gravity+matter system in one dimension. Section 3 then describes our method to extract
field theory information from the bulk in general dimensionality. By mapping standard dS physics to a system
with position dependent masses and couplings in AdS,  we can apply the well established AdS/CFT dictionary to
obtain the $d-1$ dimensional description of various UV quantities of interest in the dS/dS correspondence. In
Section 4 we present results for the heat capacity of the dual FT and the conformal anomaly. In Section 5 we
extend the dS/dS correspondence to a Black Hole/Black Hole correspondence and find that the geometry is
consistent with the properties of the field theory we uncovered before, which we then summarize in Section 6. A
discussion for time dependent backgrounds with changing cosmological constant follows in Section 7. In Section 8
we present our conclusions.

\section{Properties of the dS/dS correspondence}

\subsection{The scales of interest}

As reviewed above, the $d-1$ dimensional holographic dual of
dS$_d$ is only a pair of CFTs up to the energy scale $1/L$. On the
$d$ dimensional gravity side of the correspondence, one has a
local effective field theory description good up to the
$d$-dimensional Planck scale $M_d\gg 1/L$ (or perhaps the bulk
string scale in a stringy construction). At scales above $M_d$
quantum gravity effects become important in the bulk and one has
to appropriately UV complete the system, for example by embedding
it as a metastable dS into string theory following one of the
constructions \cite{MSS,kklt,BeckerdS,Riemann,allmoduli}. Most of
our analysis in this paper will be concerned with using the
gravity side of the correspondence to determine the behavior of
the $d-1$ theory in the range of energies $1/L < E < M_d$.

In the $d-1$ description, the Planck mass is dominated by an
induced contribution, of order $M_{d-1}^{d-3}\sim S/L^{d-3}$ where
$S\sim (M_d L)^{d-2}\sim (M_{d-1} L)^{d-3} $ is the
Gibbons-Hawking entropy of dS$_d$ and the effective species number
(central charge) of our dual low energy CFTs.

\begin{figure}
   \centerline{\psfig{figure=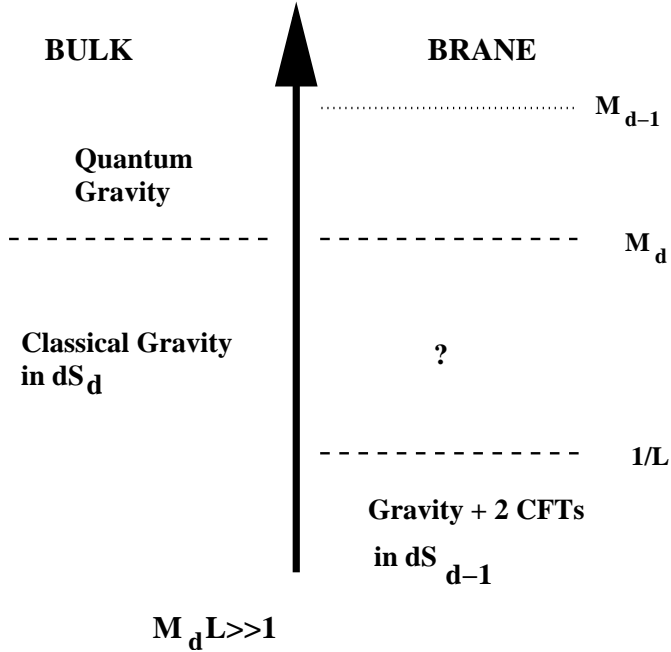,width=3.5in}}
    \caption{The hierarchy of scales.
\protect M$_{d-1}^{d-3}$ =L M$_d^{d-2}$ appears as an induced scale beyond M$_d$, the ultimate cutoff of the
theory.} \label{scales}
 \end{figure}

These scales of interest are summarized in
Fig.\ref{scales}.

Many of the calculations we perform in the next section are aimed at exploring the lower dimensional field
theory plus gravity physics in the range of energies $1/L <E<M_d$ by using bulk gravity.  The effective action
in the $d-1$ description is of the form
\beq \label{correctedaction} S_{\rm eff} = M_{d-1}^{d-3} \, \int d^{d-1} x \, \, ({\cal R} + L^2 {\cal R}^2 + L^4
{\cal R}^4 + \ldots )+\int d^{d-1}x {\cal L}_{\rm matter}. \eeq
In particular, there are extra terms in the gravitational action which arise from loops of the $S$ matter
species and are only suppressed by inverse powers of $1/L$.  In this sense, gravity becomes strong at the scale
$1/L$.  However, the interactions of matter with gravity are still suppressed; the field equations following
from the above action are of the form
\beq \label{correctedEinstein} {\cal R} + L^2 {\cal R}^2 + L^4 {\cal R}^4 + \ldots \sim {1\over M_{d-1}^{d-3}}
T_{\rm matter} \eeq
and so for matter energy scales $E$ satisfying
\beq \label{nogravityscale} E^{d-1}\ll {S\over L^{d-1}}          \eeq
the higher curvature terms in (\ref{correctedaction}),
(\ref{correctedEinstein})  are negligible.

The $d$-dimensional description makes clear that a local description of the physics is available all the way up
to the scale $M_d$, and in the regime (\ref{nogravityscale}) the interactions of matter are not strongly
corrected by gravity.  Correspondingly, one does not find $d-1$ dimensional black holes in the range $1/L < E <
M_d$.

Nonetheless, we will find that the local description in this range
of energies $1/L < E < M_d$ is softened in ways suggesting a role
for gravity.  Namely, in a precise sense we will specify, the
effective operator dimensions and the effective central charge
decrease dramatically for $E\gg 1/L$.  These effects (and the
other related effects we compute in what follows) suggest
gravitational dressing and cancellation of the conformal anomaly
in the $d-1$ theory at energies above $1/L$.

In addition, we may have couplings in the matter sector of (\ref{correctedaction}) among operators of the two
low energy CFTs at or above the scale $1/L$.  
We will see from a generalization of the calculations in
\cite{{tunnelingsmall},{tunnelingbig}} that such interactions must
be present.

We will revisit the interpretation of the physics in the range $1/L < E < M_d$ after computing the behavior of
various basic physical quantities in this regime.


\subsection{de Sitter Quantum Gravity}

For general $d$ both sides of the duality contain quantum gravity and hence need to be UV completed above the
scale $M_d$. Repeated application of the same duality can map dS$_d$ gravity in any $d$ down to 2+1, 1+1 or 0+1
dimensions, where gravity becomes non-dynamical. In particular when reducing the system down to two or one
dimension, the lower dimensional gravitational physics is a UV complete theory even above scales $M_d$.
Therefore once we established that the two theories are equivalent up to scales $M_d$, one can take the lower
dimensional system as one possible UV completion of bulk gravity! The full description of the system then is
this low dimensional gravitational system coupled to a complicated matter sector.

For concreteness let us focus to the scenario where we dualize all
the way down to 0+1 dimensions, but a similar case could be made
for 1+1 as well. What we end up with is a standard quantum
mechanical theory with Hamiltonian and Hilbert space coupled to 1d
gravity on dS$_1$. What does this mean? Since the only space-time
dimension we are left with is time, the metric is trivial. The
only sense in which we are on dS$_1$ is that we are instructed to
study the theory at a finite temperature $T=\frac{1}{2 \pi L}$.
What does it mean to couple the system to gravity? In 0+1
dimensions gravity doesn't have any dynamical degrees of freedom.
It does, however, impose a constraint \beq \label{constraint}
H={\rm const}\eeq on the Hamiltonian as a consequence of time
reparametrization invariance. Like for the 2d string worldsheet
this constraint can be imposed on the level of states.

Another consequence of (\ref{constraint}) is that there is no interesting time evolution of the physical states.
The wavefunction of the universe is fixed, there are no outside observers that could measure the wavefunction of
the universe. The best known example of such a quantum mechanical system with Hamiltonian constraint due to time
reparametrization invariance is the worldline of a relativistic point particle. For dS/dS one needs simply to
replace the $(\dot{X})^2$ term in the point particle action with a more general theory involving many degrees of
freedom.

Then what are the observables in such a constrained quantum mechanical system? Since there is no non-trivial
time evolution, the usual scattering amplitudes will not be interesting, but one can define conditional
probabilities based on chains of projection operators and decoherence functionals. This has been addressed for
example in \cite{hartle}. The original motivation for studying reparametrization invariant quantum mechanics as
a model for quantum cosmology was derived from Minisuperspace, where the space of all metrics gets truncated to
the space of all FRW like universes and the quantum gravity just becomes quantum mechanics of the scale factor.
In that case, the truncation was just a toy model; in dS/dS it is the full story derived from holography (given
the rich matter sector corresponding to the bulk geometry). This provides a new motivation to develop this
formalism as a framework for quantum cosmology.

\section{Holographic calculation in de Sitter space via AdS/CFT}

In this section, we study the linearized theory of fields -- scalars and the graviton -- in the $d$ dimensional
description.  A given bulk mode can play two roles in our system:  it sources operators in the two CFTs, and if
light enough contains an extra ``localized" $d-1$ dimensional mode independent of the degrees of freedom of the
two CFTs.  We start by reviewing the mode spectrum and applying it to calculate aspects of the communication
between the two CFT throats built into dS$_d$.  Then we move on to focus on the physics contained in each
throat, by making a conformal map to AdS in a way which allows us to read off the behavior of the effective
operator dimensions and central charge of the system in the deep UV ($E\gg 1/L$).

\subsection{Modes in dS}

In \cite{dsds} we reviewed the mode spectrum in $d-1$ dimensions descending from the dS$_d$ causal patch sliced
by dS$_{d-1}$ slices (\ref{slicing}).  At large $z/L$, there is a continuum of modes corresponding to the CFT
degrees of freedom.  In addition, there can be extra bound states in the potential supported near $z=0$ such as
the localized graviton \cite{autolocalize}. In this subsection we will determine the spectrum of such modes in
the case of bulk scalar fields of mass $M$ and comment on its interpretation in the dual.

The mass $m$ of the zero mode will depend on the bulk mass $M$ of the scalar field $X$. In order to determine
this dependence, we need to solve the analog quantum mechanics for a massive scalar field.


As discussed in \cite{dsds}, scalar fields of mass $M$ in dS$_d$ satisfy an equation of motion isomorphic to a
Schr\"odinger equation with potential
\beq \label{pot} V= \frac{(d-2)^2}{4L^2} - \left ( \frac{(d-1)^2-1}{4L^2} - M^2 \right ) \frac{1}{\cosh^2(\zol)}
\eeq
Here we separate variables as in \cite{dsds,autolocalize} and
denote by $m$ the mass of the $d-1$ particle arising from this
dimensional reduction.

Analyzing this Schrodinger problem using a mapping to an analogue supersymmetric quantum mechanics problem
yields the result
\beq m^2 = \frac{1}{2 L^2} (d-1) -M^2 -
\frac{1}{2L^2} \sqrt{ (d-1)^2 - 4 L^2 M^2}. \eeq

For $M=0$ we get a massless zero mode with $m=0$ as for the graviton. At the conformally coupled value in the
bulk, $M^2=\frac{(d-1)^2-1}{4L^2}$, the boundstate becomes marginal with $m^2=\frac{(d-2)^2}{4 L^2}$, since at this
value the $\cosh^2$ term in the potential (\ref{pot}) changes sign. For larger values of $M$ we lose the bound
state\footnote{The formula for $m^2$ however still gives a real answer
up to bulk masses of $M=\frac{d-1}{2
L}$, which is precisely the value of $M$ at which the bulk behavior changes from exponential to oscillatory
eigenfunctions ($\mu$ becomes imaginary in the language of \cite{bms}).}.

The absence of a bound state for sufficiently large masses has interesting implications.  For example, consider
the case that we have a perturbative string description in the bulk, so that amplitudes soften above the scale
$m_s$, providing a UV completion of gravity in the bulk.  In the $d-1$ description, the string mass modes will
not survive as $d-1$ particles.  Hence the $d-1$ gravity theory amplitudes are not rendered finite by stringy
physics, but instead must be UV completed via a different form of quantum gravity.  This is reminiscent of
relations between perturbative string limits and the 11 dimensional supergravity limit of M theory via
strong-weak coupling dualities; the latter is not UV completed by a tower of string states.  In our case, for
$d=4$, the holographic dual theory is $2+1$ dimensional gravity, which has been argued to be well defined on its
own \cite{Wittenloop}.

\subsubsection{Communication between throats}

In \cite{{tunnelingsmall},{tunnelingbig}} interactions between low energy field theories corresponding to
warped throats were studied via gravity-side calculations.  The interactions between operators ${\cal O}_1$
corresponding to the low energy effective theory in one throat and ${\cal O}_2$ in another were found to be
suppressed generically only by powers of the curvature radius scale $1/L$.

In our case, we can perform a similar calculation.  Let us denote the $d-1$ dimensional mass of the ``glueball"
we send across the barrier $m$.  Using the above potential barrier (\ref{pot}), the transmission probability
across the barrier works out to be
\beq \label{transmission} T(M^\prime,m^\prime)={\sinh^2(\pi m^\prime L)\over{\sinh^2(\pi m^\prime
L)+\sinh^2{\pi\over 2}\sqrt{4(M^{\prime 2} L^2-1)}}} \eeq
where $m^{\prime 2}=m^2-(d-2)^2/4L^2$ and $M^{\prime 2}=M^2-(d^2-2d)/4L^2$.

This formula exhibits the same general behavior noted in
\cite{{tunnelingbig},{tunnelingsmall}}:  the interactions between
the two throats are suppressed only by powers of energy ($m$)
divided by $1/L$.
Even after taking into account the enhancement of gravitational
interactions due to the large number of species running in the
matter loops, it seems this unsuppressed tunneling rate can not be
accounted for by gravity alone. We have to add some explicit
couplings between the 2 CFTs. Later we will argue that some
explicit couplings between the two CFTs are also required just to
get bulk fields to be continous across the UV-brane.

\subsection{Conformal mapping from dS to AdS physics}
\subsubsection{Basic Strategy}

In dS slicing the metric of dS$_d$ reads \beq ds^2_{dS_d} = \frac{L^2}{\cosh^2(z)} ( ds^2_{dS_{d-1}}+ dz^2 )
\eeq while AdS$_d$ can be written as \beq \label{ads} ds^2_{AdS_d} = \frac{L^2}{\sinh^2(z)} ( ds^2_{dS_{d-1}}+
dz^2 ) = \frac{1}{\tanh^2(z)} ds^2_{dS_{d}} \eeq
so the two
are related by a simple conformal transformation. We can use this
to map the physics in dS to dynamics in AdS, albeit with unusual
actions. Namely the conformal map yields scalars with position
dependent masses and gravity with a position dependent Newton
constant.  By applying the AdS/CFT dictionary to the resulting
system, this allows us to make a direct comparison of the UV
behavior of the $d-1$ dual of dS$_d$ to the UV behavior of a
strongly 't Hooft coupled CFT.

Note that this conformal map takes only half of our dS setup (one of the two throats) to a full AdS slice. So we
expect this analysis to capture the effective dynamics of either one of the two CFTs, with the effects of
gravity and of the other CFT folded in as they arise above the scale $1/L$.  The latter in particular will be
encoded in the boundary conditions on AdS. In the dS space fields are continuous, so after the conformal map to
AdS we get a theory on two copies of AdS where the boundary value of a given field in one copy sets the boundary
conditions for the same field in the second copy, as we will discuss in more detail later. From now on we will
mostly set $L=1$ and only restore $L$ when necessary.

The AdS/CFT dictionary relates bulk scattering amplitudes to local Greens functions in the dual CFT.  In our
case, gravity is present in the system, and hence local physics is not expected down to arbitrarily short
distance scales.  Nonetheless, formulating the computations in terms of the AdS/CFT dictionary proves useful: we
will find that because of the radially running bulk field masses and Planck mass, short distance physics is
softened in a way we can quantify.  For example, while we can express bulk amplitudes in terms of a two point
function of operators in $d-1$ dimensions, the dimensions of these operators are given by a universal finite
value in the UV for each spin of bulk field.  The specific heat, a measure of the degrees of freedom, also shuts
off in the UV in our system in contrast to AdS/CFT but in a way we can study using the above map to the AdS/CFT
dictionary.

\subsubsection{Scalar Fields}

Let us
first consider the case of a scalar field. The bulk action for a
free, massive scalar field in dS$_d$ is

\beq
\label{scalaraction} S= \int d^dx \sqrt{-g} \left (
-(\partial_{\mu} X) - (M^2 + \xi R) X^2 \right )
\eeq
where we allowed explicit mass terms as well as mass terms that
arise from a coupling to the constant background curvature.
\begin{table}[t]
\label{traf}
\begin{tabular}[t]{|lcl|}
\hline
$g_{mn}$&$ \rightarrow$&$ f^2 g_{mn} \rule{0cm}{.58cm}$ \\
$X$&$ \rightarrow$&$ f^{-\frac{d-2}{2}} X \rule{0cm}{.58cm}$ \\
$\sqg$&$ \rightarrow$&$ f^{d} \sqg  \rule{0cm}{.58cm}$ \\
$ - \sqg(\partial X)^2$ &$\rightarrow$&$ -\sqg(\partial X)^2 -
\sqg \frac{(d-2)}{2} X^2 (\nabla^2 \omega) -
 \sqg \frac{(d-2)^2}{4} X^2
 (\nabla
 \omega)^2  \rule{0cm}{.58cm}$ \\
$\sqg R$&$ \rightarrow$&$ f^{d-2} \sqg  (R -
2 (d-1) (\nabla^2 \omega) - (d-2)(d-1)(\nabla \omega)^2 )
  \rule{0cm}{.58cm}$ \\
$ - \sqg \xi R X^2 $&$\rightarrow$&$ -
\sqg \xi R X^2 +2  \sqg \xi (d-1) (\nabla^2 \omega) +
 \sqg \xi  (d-2)(d-1)(\nabla \omega)^2
 \rule{0cm}{.58cm}$ \\
$- 2 \sqg \Lambda $&$ \rightarrow$&$- 2 f^{d} \sqg \Lambda
  \rule{0cm}{.58cm}$ \\
\hline
\end{tabular}
\caption{Transformations under conformal rescaling; $\omega=\log f$}
\end{table}
\noindent Table (\ref{traf}) summarizes the transformation
properties of the various terms in the action under a conformal
rescaling with a general function $f$, where $\omega=\log f$. We
see that the scalar kinetic term reproduces itself together with
some mass like terms. Also the explicit mass terms change. So a
scalar of mass $M$ in dS maps again to a scalar field in AdS, but
with a different mass $M_{\rm total}$. Since the conformal factor $f$
depends on the $z$-coordinate, these are position dependent masses
in AdS. In order to determine the dimension of the dual operator
one only needs the UV value of that mass, that is the value it
takes close to the boundary. Of course the full function
$M_{\rm total}(z)$ will encode interesting information of how the
dimensions of operators evolve as we go from the UV to the IR. For
our case we need
\beq
ds^2_{dS_d} = \tanh^2(z) ds^2_{AdS_d}, \; \; \; \, \, \,\, \, f=\tanh(z).
\eeq
With this we get for the AdS mass:
\beq M_{\rm total}^2 = \tanh^2(\zol) \left ( M^2
+  \xi d (d-1) \right )  - \frac{d(d-2)}{4}
\left ( 1+\tanh^2(\zol) \right ).
\eeq
where we used $R=-d(d-1)$ for the AdS$_d$.
In the UV ($z=0$) the original dS mass term M scales
to zero, as does the original $\xi$ term. Instead we get the universal result
\beq M_{\rm total}^2 = - \frac{d (d-2)}{4}
    \, \, \, \, \,\,\,    \mbox{ for all }\xi, \, M.
\eeq
That is we get a conformally coupled scalar in AdS, independent of
what values of the parameters $M$ and $\xi$ we started with in dS!
The corresponding UV dimension of the dual operator is
\beq \Delta_{O} = \left \{ \begin{array}{l} \frac{d}{2} \\ \frac{d}{2}-1
\end{array} \right. .
\eeq
This ensures that the $< OO >$ two point function for the second
choice reduces to the usual $\frac{1}{|x|^{d-2}}$ behavior of a
scalar field in $d$ dimensions.

\begin{figure}
   \centerline{\psfig{figure=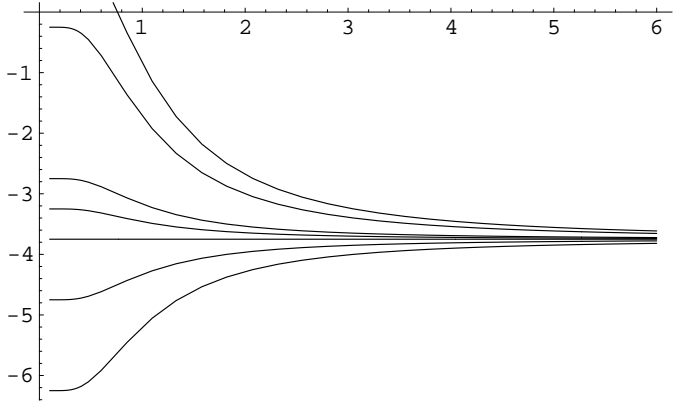,width=3.5in}}
    \caption{
The position dependent AdS masses in $d=5$ for various values of
$M$ and $\xi$.
}
\label{universal}
 \end{figure}

\subsubsection{Gravity}

Let us repeat the same analysis for gravity. We start with
the Einstein-Hilbert action in dS
\beq
\label{eh} S = M_d^{d-2} \int d^dx \sqrt{-g} ( R - 2 \Lambda)
\eeq
where $\Lambda= \frac{1}{2} (d-1) (d-2)$. From table (\ref{traf})
we see that similar to the bulk scalar mass $M$ the original dS
cosmological constant $\Lambda$ scales away. A big difference to
the scalar field case this time is, however, that the
gravitational kinetic term does {\it not} reproduce itself,
instead we get a new gravitational action in AdS
\begin{eqnarray}
\label{rungn}
\nonumber
S &=& \int d^d x (M_d f)^{d-2} \sqrt{-g}
\left ( R - 2(d-1) ( \nabla^2 \omega) - (d-2)(d-1) (\nabla \omega)^2
- 2 f^2 \Lambda \right ) \\
& =_{ibp}&
\int d^d x \sqrt{-g} (M_d f)^{d-2} \left ( R + (d-2)(d-1) (\nabla \omega)^2
- 2 f^2 \Lambda \right ).
\end{eqnarray}
We see that the dS/dS graviton corresponds to a varying Newton's
constant in AdS, $M_d(z) = f(z) M_d$. As we approach the boundary
the gravitational coupling increases. This allows for the
localized graviton. The cosmological term is no longer constant,
but involves covariant derivatives of $\omega$. We will study
solutions to the full non-linear equations that follow from
(\ref{rungn}) momentarily. To find the dimension of the operator
dual to the graviton let us first consider linearized
fluctuations. One expects the transverse traceless fluctuations of
the graviton to satisfy the same equations as a massless scalar
field. To read off the behavior of the dual FT operator, we only
need the small $z$ properties, where $\tanh(z) \sim z$. So our
graviton with position dependent Planck mass should be equivalent
to studying an auxiliary action
\beq
S = \int d^d x z^{d-2} \sqrt{-g} \partial_{\mu} \Phi \partial^{\mu} \Phi.
\eeq
The equations of motion are
\beq
\frac{1}{\sqrt{-g}} z^{2-d} \partial_{\mu} z^{d-2}
\sqrt{-g} g^{\mu \nu} \partial_{\nu}  \Phi = \partial_{\mu}
\partial^{\mu}  \Phi =0.
\eeq
Close to the boundary the graviton looks like a flat space graviton!
In particular, the possible boundary behaviors are $z^0$ and $z^1$, as
opposed to $z^0$ and $z^4$ for ordinary AdS gravity.
This seems to yield two possibilities of the dimension of the dual operator,
$d-1$ and $d-2$. Of course we expect the graviton to couple to the
energy-momentum tensor with dimension $d-1$.

\section{The UV structure of the holographic dual}
\subsection{Finite temperature}
\subsubsection{Equations of Motion}

The most pressing question about the UV
behavior of the dual theory is to what extent it is described by a
local field theory coupled to gravity. One possible approach is to study the system at finite temperature. In
particular the heat capacity of the system tells us how the degrees of freedom grow at high energies. A
vanishing heat capacity, like found for the non-extremal NS5 branes, signals that the dual theory, in that case
a little string theory, would certainly not be a local QFT. Alternatively, in even dimension one can calculate
the Weyl anomaly following \cite{hensken}, which we will do in the next section. For both approaches one needs
to construct solutions to the bulk equations of motion. In order to find those we need to first derive the
equations of motions that follow from the action (\ref{rungn}):

\begin{eqnarray}
\label{eom}
\nonumber
0&=&(R_{\mu \nu} - \frac{1}{2} g_{\mu \nu} R) - f^{2-d}
\left \{ \nabla_{\mu} \nabla_{\nu} f^{d-2} - g_{\mu \nu}
\nabla^2 f^{d-2} \right \} + \\
&&+ (d-2)(d-1) \left \{ (\nabla_{\mu} \omega) (\nabla_{\nu} \omega) -
\frac{1}{2} g_{\mu \nu} (\nabla \omega)^2 \right \} +  g_{\mu \nu} \Lambda f^2
\end{eqnarray}
As a first check it is straight forward to verify
that for our choice of $f$ and $\Lambda$,
$f=\tanh(z)$ and $\Lambda=\frac{1}{2} (d-1) (d-2)$,
AdS$_d$ written in the coordinate system (\ref{ads}) indeed solves
the equations of motion (\ref{eom}).

\subsubsection{Black Brane Solutions}

In order to study the thermodynamic properties of the theory
we would like to find the black brane solutions to (\ref{eom}).
To map out the UV properties it is sufficient to study the
somewhat simpler system with $f=z$ and $\Lambda=0$. What this corresponds
to is the conformal map of the $z>0$ half of Minkowski
space M$_d$ to AdS$_d$,
\beq ds^2_{AdS_d} = \frac{1}{z^2} \left ( - dt^2 + d \vec{x}^2 + dz^2
\right )=
\frac{1}{z^2} ds^2_{M_d}
\eeq
For small $z$, $\tanh(z) \sim z$; in addition as $z \rightarrow 0$
the curvature on the dS$_{d-1}$ slice
can be neglected. So in the UV the thermodynamic
properties of dS/dS can be studied by looking for black brane solutions
of (\ref{eom}) with $f=z$ and $\Lambda=0$.

After changing coordinates to $r=1/z$, it is again straight forward to check
that AdS$_d$ written as
\beq
\label{ansatz}
ds^2 = - h(r) dt^2 + \frac{dr^2}{h(r)} + r^2 d \vec{x}^2
\eeq
with
\beq
h(r) = r^2
\eeq
solves the equations of motion (\ref{eom}). With
the same ansatz (\ref{ansatz}) the most general solution for $h(r)$ is
\beq
\label{blackbrane}
h(r) = r^2 - \mu r
\eeq
which describes a black brane in asymptotically AdS space with a horizon
at
\beq r_H= \mu.
\eeq
The temperature of this black brane can be obtained by Wick rotating to
Euclidean space.
To avoid a conical singularity, $t_E$ must have periodicity
\beq \beta= \frac{1}{T} = \frac{ 4 \pi}{\mu}.\eeq
To determine the free energy $F$ we can calculate the difference of euclidean
on-shell gravitational actions
\beq
I = \beta F = -  \frac{1}{16 \pi G_N} \int d^d x \sqrt{g} f(z)^{d-2} \left (R
+ (d-2)(d-1) (\nabla \omega)^2 - 2 f^2 \Lambda \right ) + I_{\rm GH}
\eeq
between our solution and thermally compactified AdS as the reference
spacetime with $F=0$. In the usual Hawking Page analysis \cite{hp,wthermal}
the on-shell value of the integrand is $-2 (d-1) \sqrt{g}$ so that
$16 \pi G_N I$ just becomes $2(d-1)$ times the proper volume of spacetime.
Also the Gibbons Hawking term $I_{GH}$ in that case does not contribute,
since the term in $h(r)$ proportional to $\mu$ falls of to rapidly with $r$.
Plugging back (\ref{blackbrane}) into the action we find that in our case
the integrand is just $-2 (d-1)$, since the $\sqrt{g}$ factor cancels against
the position dependence of $G_N$. So instead of calculating the volume,
in our case the action just measures the coordinate volume. Matching the
periodicity $\beta'$ of thermal AdS to the period of $t_E$ at a fixed, large
value $R$,
\beq \beta' R = \beta \sqrt{R^2 - \mu R}\, \, \, \,\, \, \Rightarrow
\, \, \, \,\,\, \beta' \approx \beta
\left (1 - \frac{1}{2} \frac{\mu}{R} \right ),
\eeq
and compactifying the $\vec{x}$ coordinates on a volume $V$
we find the bulk contribution to the free energy to be
\beq
16 \pi G_N \beta F_{\rm bulk} = 2 (d-1) V \left ( \beta (R-r_h) - \beta' R \right )=
-(d-1) V \beta r_H = - 4 \pi (d-1) V.
\eeq
In addition we also have a contribution from the Gibbons Hawking term.
For a metric of the form (\ref{ansatz}) the trace of the extrinsic curvature
$\Theta_{ab}$ of a slice at $r=R$ is
\beq
\Theta(r=R) = - h^{1/2} \left ( \frac{h'}{2h} + \frac{d-2}{r} \right ).
\eeq
With this we get for the black brane or thermal AdS background
\beq
16  \pi G_N I_{\rm GH} = f(R)^{d-2} \int d^{d-1}x \sqrt{g_I} \Theta = \int d^{d-1}x
\left ( (2d-3) \mu - 2 (d-1) R \right )
\eeq
where again the position dependence of Newton's constant with
$f(r)=1/r$ cancelled against the one in the square root of the
determinant of the induced metric $\sqrt{g_I}$. Finally we get for
the action difference
\beq
16 \pi G_N \beta F_{\rm GH} = V  \beta
 \left (  (2d-3) \mu - 2(d-1) R \right
) +
V \beta' 2 (d-1) R  = 4 \pi (d-2) V
\eeq
Putting bulk and GH contribution together we finally obtain:
\beq \beta F = -\frac{V}{4 G_N}.
\eeq
$F$ is negative, so for all values of $\mu$ the black
brane is thermodynamically preferred over thermal AdS. However,
$\beta F$ is independent of $\beta$, so the internal energy
$E=\frac{\partial I}{\partial \beta}$ vanishes, and so does
the heat capacity. The system has basically no degrees of freedom,
the internal energy is independent of the temperature! Last but
not least, one can calculate the entropy
\beq
S= \beta E - I = - I = \frac{V}{4 G_N},
\eeq
which is precisely the usual quarter of the
horizon area $V r^{d-2}$ in units of the local Newton's constant,
$G_N(r)=G_N r^{d-2}$.
While the analysis of the black brane for the $f=z$ theory shows that
the heat capacity of the dS/dS system is approaching zero in the
far UV, in order to determine the approach to zero one would have to do
a similar analysis for the black branes in the $f=\tanh(z)$ theory. This
is beyond the scope of the current analysis.

To close this section let us try to understand the physics of the
black branes we found by undoing the conformal map that took us from Minkowski
space to AdS. Instead of being solutions to an unconventional
gravitational action in an AdS background they should then correspond simply
to solutions of vacuum Einstein gravity.
The black brane metric is conformal
to
\beq
ds^2 = - \frac{h(r)}{r^2} dt^2 + \frac{dr^2}{r^2 h(r)} + d \vec{x}^2
\eeq
which in terms of
\beq
\rho = \frac{2 (r-\mu)}{\mu \sqrt{h(r)}}
\eeq
reads
\beq
ds^2 = -\frac{\mu^2 \rho^2}{4} dt^2 + d \rho^2 + d \vec{x}^2.
\eeq
In the original conformal frame all black brane solutions
are just Rindler spaces with different time coordinates!
Many of the thermodynamic properties of the black branes follow
directly from this identification, but their interpretation in terms
of a dual CFT required us to transform to AdS.

\subsection{Conformal Anomaly}

AdS/CFT instructs us to evaluate the bulk action on a given solution
in order to calculate the boundary partition function for
a given boundary metric. This quantity has divergences due to the infinite
volume of AdS. In a coordinate system where the boundary sits at $z=0$
the divergences will typically go like powers of $z$. These divergences
map one-to-one to UV divergences in the field theory and can be dealt
with in the same way: by adding local counterterms on the $z=\epsilon$ slice
before taking $\epsilon \rightarrow 0$. In order to preserve
diffeomorphism invariance these counterterms should be constructed
from curvature invariants made out of the induced metric on the slice.
For details on the procedure see \cite{hensken}. However in
even boundary dimensions (odd bulk dimension $d$)
there are in addition $\log(z)$ terms and they represent
the conformal anomaly.

For the standard Einstein Hilbert action on the
dS$_{d-1}$ sliced AdS$_d$ background (\ref{ads})
the on-shell value of the integrand is $-2 (d-1) \sqrt{g}$, as
reviewed above in the black brane context. Hence
\beq
S_{\rm on-shell} = \frac{- 2(d-1)}{16 \pi G_N} \int
\frac{dz}{\sinh^d(z)} \sim  \cosh(z)\; _2F_1(\frac{1}{2},
\frac{d+1}{2},\frac{3}{2},\cosh^2(z)).
\eeq
Expanding the rhs in powers of $z$ around $z=0$ one finds that in
odd dimension the only divergent terms that appear are powers of
$z$, while in even dimension there is indeed in addition a
logarithmic term. In particular in $d=3$,
5, 7 we obtain
\begin{eqnarray}
\int \frac{dz}{\sinh^3(z)} & =&  -\frac{1}{2z^2} -
\frac{\log(z)}{2}  + {\cal O}(z^0)\\
\int \frac{dz}{\sinh^5(z)} & =&  -\frac{1}{4z^4} +\frac{5}{12 z^2}
+ \frac{3}{8} \log(z)
+ {\cal O}(z^0)\\
\int \frac{dz}{\sinh^7(z)} & =&  -\frac{1}{6z^6} + \frac{7}{24z^2}
- \frac{259}{720 z^2} - \frac{5}{16} \log(z)  + {\cal O}(z^0)
\end{eqnarray}
The $\log$ terms give the conformal anomaly evaluated on dS$_2$,
dS$_4$ and dS$_6$ respectively. In the conventions of
\cite{hensken} we want to write the anomaly as
\beq {\cal A} =-\frac{2}{16 \pi G_N} a_{d-1}. \eeq
where $-\frac{2}{16 \pi G_N} a_{d-1}$ is the coefficient of the $\log(z)$
term in the on-shell
action\footnote{The extra factor of 2 comes about from the
difference between our $z$ and their $\rho$ coordinate. What is
important is how the action transforms under a conformal
transformation and that is captured by ${\cal A}$. We put the factor
of 2 into the definition of $a_{d-1}$ in order to agree with \cite{hensken}
already on the $a$'s}. So we get for dS$_{d-1}$ backgrounds
\begin{eqnarray}
a_2&=& 1  \\
a_4&=& -\frac{3}{2}  \\
a_6&=& \frac{15}{4}
\end{eqnarray}
In 2d this uniquely fixes the anomaly.
\beq {\cal A} = -\frac{c}{24 \pi} R \eeq
and one can just read off $c=\frac{3}{2 G_N}$. In 4d the conformal
anomaly is given in terms of two numbers, the coefficient $a$ of
$C^2=R_{\mu \nu \rho \sigma} R^{\mu \nu \rho \sigma} - 2 R_{\mu
\nu} R^{\mu \nu} + R^2/3$ and the coefficient $-c$ of $R_{\mu \nu
\rho \sigma} R^{\mu \nu \rho \sigma} - 4R_{\mu \nu} R^{\mu \nu}
+R^2$. In order to determine both $a$ and $c$ we need to know the
bulk solution for one more boundary metric. A trivial extension is
to consider the ``black-string'' type metric, where the dS$_{d-1}$
space on the slice gets replaced with the dS$_{d-1}$ Schwarzschild
black hole. It is well known that this metric solves the equation
of motions with the same value of the on-shell action, since the
$d$-dim $R$ only depends on the $(d-1)$-dim $R_{\mu \nu}$ on the
slice, not on $R_{\mu \nu \rho \sigma}$. From this we immediately
learn that $a=c$, since the Riemann$^2$ terms have to cancel from
the anomaly. With this choice our value for $a_4$ agrees perfectly
with the one found by \cite{hensken} when evaluated on a dS or SdS
background. In 6d three tensor structures show up and our crude
methods of just plugging in two backgrounds for which we happen to
know the full solution into the action only give us two linear
combinations of those. Still, our value of $a_6$ is again in
perfect agreement with \cite{hensken} evaluated on dS$_6$.

Now let us repeat the same exercise for the gravitational action
(\ref{rungn}) with position dependent $G_N$. The position
dependence of $M_d$ is $M_d(z)=\tanh(z) M_d$ which gives an extra
factor of $\tanh^{d-2}(z)$. Up to terms that remain finite as $z
\rightarrow 0$
\beq
S_{\rm on-shell} = \int  \tanh^{d-2}(z) \frac{dz}{\sinh^d(z)} = \int
\frac{1}{\sinh^2(z)} \frac{dz}{\cosh^d(z)} = \frac{-1}{z} + {\cal O}(z).
\eeq
For all $d$ the only divergent term is a universal $-\frac{1}{z}$ and
there are no logarithms. The conformal anomaly vanishes\footnote{A similar 
analysis has been performed in the context of the dS/CFT correspondence
of \cite{dscft} in \cite{citeme}. In that case the putative dual does
not involve gravity and the conformal anomaly doesn't vanish.}

One possible interpretation is that lower dimensional gravity
screens the central charge to be zero, just like is well known
from 2d gravity on string theory worldsheets. In this scenario one
does not even need a conformal field theory beyond scales $1/L$
since the gravitational dressing will also make any FT a CFT.

In
order to explore this a little further let us use the same methods
to evaluate the conformal anomaly for a smooth RS type domain wall
as constructed in \cite{dfgk} with dS slices. The metric can be
written in the form
\beq
ds^2 = e^{2 A(z)} \left ( ds^2_{dS_{d-1}} + dz^2 \right ).
\eeq
and so the conformal map can be performed just as in the dS$_d$
case with $f=\frac{\sinh(z)}{e^A}$ instead of $f=\tanh(z)$. A
graviton localizes at the center of the brane where $A'=0$.
We can call this place $z=0$. As long as
$e^A(0)$ is finite, the gravitational part of the on-shell action
still evaluates to $\frac{-1}{z} + {\cal O}(z)$. The only part of
$f^{d-2}$ that was essential in this result was that it vanishes
as $z^{d-2}$ for $z \rightarrow$ 0 in order to lower the degree of
divergence. The $\cosh(z)$ only matters when looking at the finite
terms.
To
evaluate the full conformal anomaly, one needs to calculate the
on-shell action for all fields. The fat walls of \cite{dfgk} are
built from bulk scalars. We already worked out the conformal
transformation properties of (\ref{scalaraction}) before. While
the scalar potential terms become irrelevant once transformed to
AdS, the kinetic term remains unchanged. So we get a contribution
to the on-shell action
\beq
S_{\rm on-shell}^{\rm scalar} = \int d z \sqrt{-g} (\Phi')^2 =
\int d z \frac{(\Phi')^2}{\sinh^d(z)}
\eeq
After integration by parts and use of the equations of motion one
sees that the the bulk contribution to the conformal anomaly is
proportional to $\Box_{d-1} \Phi$ and hence vanishes for field
configuration that are constant on the slice, as we want in order
to preserve the symmetry on the slice. The boundary terms one
generates when integrating by parts do not vanish, in fact they
are singular. However since they only involve powers of the
$\sinh$ and not its integral, all divergences are power
divergences that can be cancelled by local counterterms. No log
terms arise. This result was first obtained
in \cite{skensol} who analyzed the conformal
anomalies in general backgrounds with scalars turned on. This supports our
conclusion that the zero central charge is due to gravitational
dressing cancelling the matter central charge.

In the same spirit the universal UV dimension of the scalar fields can be understood as gravitational dressing.
Naively one would think that gravitational dressing should bring the operator dimension to $d-1$ so that one can
add it to the action. But the $(d-1)/2 \pm 1/2$ we find is consistent as long as we only add products of the
form $O^1_{d/2} O^2_{d/2-1}$ to the action\footnote{ We note also that in our reparametrization of dS slicing of
dS space the symmetry which is manifest is $SO(1,d-1)\times Z_2$ where the $Z_2$ factor exchanges two CFTs.
Therefore we should add the same operator to the action with 1 and 2  interchanged.}, where the subscripts label
the dimension of the operator and the superscripts the two CFTs. We know that the coupling of the two CFTs has
to be achieved via its boundary interactions. Continuity across the UV brane in the original dS space means that
the value of the field at the boundary in one AdS (dual to CFT 1) appears as a boundary condition in the second
AdS (dual to CFT 2). The discussion of multi-trace operators in \cite{wittenmulti,berkooz} uses precisely 
the product
operator ${\cal O} =O^1_{d/2} O^2_{d/2-1}$ in order to achieve boundary conditions of the type we want, at least
in a folded version of our duality: instead of one scalar field living in 2 copies of AdS there one has 2
decoupled scalar fields in one copy of AdS. Since we are dealing with gravity in addition to scalar fields, for
us the 2 copies of AdS are more appropriate in order to avoid having 2 gravitons living in the same space. The
effect of adding ${\cal O}$ to the action is to set the leading and the subleading behavior of the scalar field
on one AdS equal to the leading and subleading term in the other AdS respectively.\footnote{As discussed in
\cite{doubletrace}, the addition of a single product of operators would lead to extreme nonlocality in the {\it
internal} dimensions $X$ of the $AdS\times X$ solutions in AdS/CFT because each operator corresponds to a
particular Kaluza-Klein mode in the internal space and hence each factor in the product involves an integral
over the internal space. In order to avoid that here, we can add an appropriate infinite linear combination of
product operators corresponding to a local boundary condition in all dimensions.}

\section{The Black Hole/Black Hole correspondence}
\label{bblackhole}

For further support for our picture of the FT
that governs the scales $1/L < E < M_d$ we turn to the study of
brane world black holes. According to \cite{kaloper}, the
classical solution describing a brane world black hole from the
bulk point of view encodes the quantum corrections to the lower
dimensional black hole solution due to the (C)FT.

Let us briefly review the arguments of \cite{kaloper} in the dS/dS
context. For Hawking radiation to not backreact, one needs the
Hawking temperature to be much less than the black hole mass M:
\beq
T \sim \frac{1}{r_H} << M, \, \, \, \, \, \,\,   \mbox{ where } \,
\, \, \, r_H^{d-3} = \frac{M}{M_d^{d-2}}
\eeq
Hence one need $M >> M_d$. The black hole has
to be much heavier than Planck scale. Since $M_d << M_{d-1}$, there is
an interesting window
\beq
M_{d-1} >> M >> M_d
\eeq
where the classical $d$-dimensional bulk solution remains
uncorrected, while the brane world black hole has sub-planckian
mass and hence is dominated by the backreaction of its own Hawking
radiation.
By constructing the classical bulk solution we therefore get
information on the one point function of the
stress energy tensor of the (C)FT.
In particular, in the
absence of a conformal anomaly, which we have shown to vanish for
the dS/dS system in the last section, the Hawking radiation gives
rise to a traceless energy momentum tensor.

While for the dS/dS correspondence it did not matter which
observer and hence which
dS$_{d-1}$ slicing of dS$_d$ to pick - they are all related by
dS$_d$ transformations - for the black hole it is important. One choice
is to study a black hole "behind" the cosmological horizon. In this case
we expect the CFT to have a finite temperature which gets quickly diluted
due to the dS expansion. For our purposes the more interesting slicing
is the one where we take the bulk black hole intersecting the built
in UV brane. In this case the standard Schwarzschild dS (SdS) bulk
black hole becomes the quantum corrected brane world black hole.

\begin{figure}
   \centerline{\psfig{figure=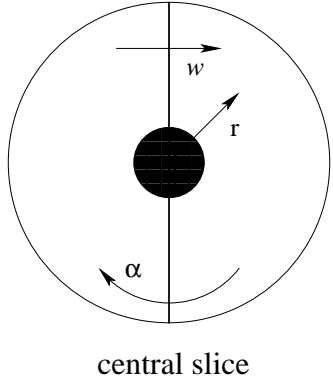,width=1.7in}}
    \caption{The bulk black hole appears as a quantum corrected brane black hole.}
\label{BHslice}
 \end{figure}

To study its properties,
let's recall some of the basic aspects of the dS/dS correspondence
\cite{dsds}. For pure dS we were using two coordinate systems:
\beq
ds^2 = - \cos^2(\theta) dt^2 + d \theta^2 +
 \sin^2(\theta)\left(d \alpha^2 + \sin^2(\alpha) d \Omega^2\right) \, \, \, \,    \mbox{
 (static)}
\eeq
and
\beq ds^2= \sin^2(w) \left( -\cos^2(\psi) dt^2 + d \psi^2 + \sin^2(\psi)
d \Omega^2\right) + dw^2 \, \, \, \,  \mbox{ (dS sliced) }
\eeq
The change of variables is simply
\begin{eqnarray}
 \cos(\theta) &=& \sin(w) \cos(\psi) \\
 \cos(\alpha) &=&
\frac{\cos(w)}{\sin(\theta(w,\psi))}
\end{eqnarray}
Now the bulk SdS black hole is easy to write
down in the static slicing ($r=\sin(\theta)$)
\beq
ds^2 =-h(r) dt^2 + \frac{dr^2}{h(r)} + r^2 \left (d \alpha^2 +
\sin^2(\alpha) d \Omega_{d-2}^2 \right)
\eeq
where
\beq
h(r) = 1- r^2 - \frac{2 \mu}{r^{d-3}}
\eeq
Even though finding the analog of the $w$-$\psi$ slicing is
difficult, it is easy to see that the $w=\pi/2$ slice (the built
in UV brane) is still mapped to $\alpha=\pi/2$. This is
fortunately all we need to know. So the metric on the brane is:
\beq
ds^2 =-h(r) dt^2 + \frac{dr^2}{h(r)} + r^2 d \Omega_{d-3}^2.
\eeq
At first glance this looks like SdS on the brane again. But note
that it has the wrong $h(r)$! This is still the $h(r)$ of d
dimensions with its $\frac{2 \mu}{r^{d-3}}$ instead of the
$\frac{2 \mu}{r^{d-4}}$ we would need in the appropriate $h(r)$
for a $(d-1)$ dimensional SdS black hole. Given the metric, one
can work out what is the stress tensor that supports it, e.g for
\begin{eqnarray}
7\rightarrow6 &:& T^{\mu}_{\nu} =\frac{5}{L^2} \mbox{
diag}(1,1,1,1,1,1) +
\frac{\mu}{r^6} \mbox{ diag}(4,4,-2,-2,-2,-2) \\
\nonumber 6\rightarrow5 &:& T^{\mu}_{\nu} =\frac{4}{L^2}
\mbox{diag}(1,1,1,1,1)  +
\frac{\mu}{r^5} \mbox{ diag}(3,3,-2,-2,-2) \\
\nonumber 5\rightarrow4 &:& T^{\mu}_{\nu} =\frac{3}{L^2} \mbox{ diag}(1,1,1,1)
+
\frac{\mu}{r^4} \mbox{ diag}(2,2,-2,-2) \\
\nonumber 4\rightarrow3 &:&
T^{\mu}_{\nu} =\frac{2}{L^2} \mbox{ diag}(1,1,1)  +
\frac{\mu}{r^3}
\mbox{ diag}(1,1,-2)\\
\nonumber d\rightarrow d-1 &:& T^{\mu}_{\nu}
=\frac{d-2}{L^2} \mbox{ diag}(1,\ldots,1) + \frac{\mu}{r^{d-1}}
\mbox{ diag}(d-3,d-3,-2,\ldots,-2)
\end{eqnarray}
The important conclusion here: $T_{\rm Hawking}$ is traceless! It is
also straight forward to check that the $T^{\mu \nu}_{\rm Hawking}$ we
obtain this way is covariantly conserved, $\nabla_{\mu} T^{\mu
\nu}_{\rm Hawking} =0 $. Using Mathematica we have confirmed that
$T_{\rm Hawking}=0$ holds as well in Kerr-de-Sitter backgrounds, as we
have explicitly done for the cases of Kerr-dS$-5$ and Kerr-dS$_4$
\cite{kerrds}. They work the same way as their Schwarzschild
cousins, on the central slice one gets something that almost looks
like the lower dimensional Kerr black hole except for having the
wrong power of $r$ coming with the mass parameter in the metric
function; the angular momentum part is completely standard.

\section{Synthesis}

Let us summarize our results for the regime $1/L < E < M_d$. Using the bulk description, which is still well
approximated by classical gravity and field theory in this regime, we found that the holographic dual has
effectively zero central charge (as evinced by vanishing conformal anomaly and asymptotic heat capacity) and
scale dependent operator dimensions that in the UV tend to a universal value. The quantum stress tensor of the
dual is traceless.

All these phenomena can still be interpreted as two conformal field theories coupled to gravity. As is well
known from 2d gravity coupled to 2d CFTs, the dynamics of the conformal factor of the metric effectively cancels
out any scale dependence of the matter sector and cancels the conformal anomaly due to the non-vanishing central
charge of the matter sector. In addition, operators get gravitationally dressed since they are multiplied by
powers of the scale factor. The dressed operators have a universal scaling dimension.

Continuity of the bulk fields requires that in addition we add direct double trace couplings between the two
CFTs. The presence of some additional couplings between the two CFTs is also required for the tunneling
transmission coefficient (\ref{transmission}).

Although these results are qualitatively similar to those from
solvable $2d$ gravity models, the details are different (for
example the scaling dimensions of operators on each side are
dressed not to dimension $d-1$ but to dimension $d/2,(d-2)/2$.  It
would be very interesting to interpret our results in detail using
directly the holographic dual theory (\ref{correctedaction})
including the interactions deduced from our computations here.

\section{Changing Cosmological Constant and RG flow}

The dS/dS correspondence shows that $dS_d$ is holographically equivalent to a system which at low energies
reduces to a pair of conformal field theories in $d-1$ dimensions, coupled to $d-1$ dimensional gravity.

In this section, we generalize this correspondence to situations like inflation and landscape decays where the
cosmological term decreases in time. We find that this process in $d$ dimensions is dual to a time dependent RG
flow in $d-1$ dimensions, in which new light degrees of freedom come down as time evolves forward (realizing
something similar to the idea in \cite{andy}, but in a Lorentzian signature holographic correspondence). We also
discuss the dual description of the formation of density perturbations, which involves the non-equilibrium
statistical physics of the dual theory in the presence of the increasing number of light species.

The discussion in this section is qualitative, but this is sufficient to establish the effect of interest.

\subsection{Inflation and RG}

The dS/dS correspondence arises from foliating dS$_d$ by dS$_{d-1}$ slices
\beq \label{slicingagain}  {ds^2_{dS_d}={1\over
\cosh^2({z\over L})}(ds^2_{dS_{d-1}}+dz^2) } \eeq

There is a convention in the description of the scales in this slicing, which is inconsequential in the case of
a fixed cosmological constant \cite{dsds} but which will play a role in our later discussion.  Namely, we
identified the curvature radius and inverse cutoff of the $d-1$ theory with the curvature radius $L_d$ of
$dS_d$. However, we could rescale the radius of the $d-1$ slices to some independent value $L_{d-1}$, as long as
we maintain the relation that this radius scales like the inverse cutoff of the $d-1$ theory, $M_{UV}\sim
1/L_{d-1}$ in the sense discussed above.  In this duality the number of effective species emerges as the
Gibbons-Hawking entropy $N_{\rm species}\sim S= L^2 M_4^2$.

\begin{figure}
   \centerline{\psfig{figure=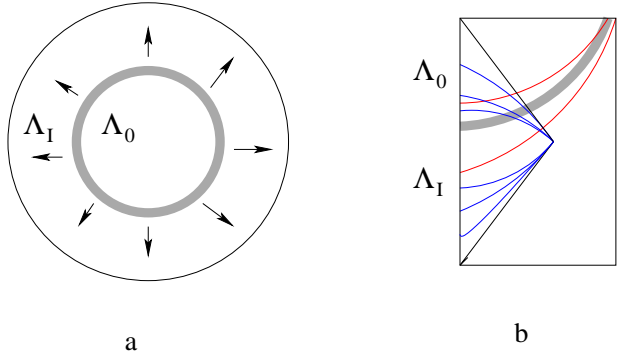,width=3.2in}}
    \caption{Penrose diagram and spatial slice of inflationary spacetime in the causal patch.}
\label{pen}
 \end{figure}

Now let us generalize this to inflation.  In the 4-dimensional gravity theory, consider specifically a spatially
flat slicing of the inflationary evolution
\beq \label{metric} {ds^2=-dt^2+a(t)^2 d\vec x^2} \eeq
with $a(t)\sim e^{H_It}$ for early times and $a(t)\sim e^{H_0 t}$
for late times, with $H_0 \ll H_I$. This is depicted in Fig. \ref{pen} (a),
with constant $t$ slices in grey, and corresponds to a tall
Penrose diagram as in \cite{tallPenrose,myers}. The Hubble
parameter varies with time $t$ from an initial value $H_I$ to a
final value $H_0$.

If we project this onto the causal patch, as discussed in
\cite{tallPenrose}, we obtain a dynamics as depicted in Fig. \ref{pen} (b),
with constant time slices for our $3d$ observer indicated in blue.
Namely, the change in cosmological constant corresponds to a
growing bubble of a phase $H_0$ inside a phase of $H_I$.  The
bubble wall may be thin or thick, depending on how fast the roll
is from one de Sitter like phase to the other.  A similar picture
describes the result of non-perturbative bubble nucleation, in the
case where the $H_I$ phase is metastable.

As discussed above, in the dS/dS correspondence the number of species in each phase is $N_i\sim M_4^2/H_i^2 $.
The radial direction $z$ corresponds to energy scale (with large $|z|$ closer to the horizon, infrared in the
$3d$ description).  Putting these features together with the inflationary dynamics, we see looking radially that
the bulk of the inflationary dynamics corresponds to a time dependent RG flow in the 3-dimensional description.
Namely, as we move from $z=0$ to $z=\infty$, toward the infrared, we cross the bubble wall and the number of
species decreases from $N_0$ to $N_I$.  This describes a renormalization group flow in which the number of
species decreases toward the IR, via running of masses and couplings in the dual description.  In addition,
these running masses and couplings are time dependent, in such a way that at fixed scale new light $H_0$ phase
degrees of freedom are brought down as time evolves forward.

It is also true that the transition appears as an expanding bubble in the $3d$ spacetime as well, though this
feature would disappear in the reduction of the correspondence to the step $d=2 \rightarrow d=1$.  Furthermore,
we can now make use of the ambiguity discussed above in the scaling of the dS$_{d-1}$ slices.  Namely, by using
the ambiguity in the scale $L_3$, we can maintain the same geometry on both sides of the bubble wall, thus
relegating the entire effect of the transition to the question of the number of light species in the different
phases.

Thus we arrive at the simple statement that inflation is holographically dual to a time dependent RG flow
process in which new light species are brought down in time.

As discussed above, the original dS/dS correspondence is limited to a low energy effective field theory
statement in $d-1$.  The inflationary physics we are interested in here has largely to do with this regime,
roughly because it is a theory of large scale physics with a resulting structure formation process which
involves largely horizon physics. Once
the bubble reaches the $|z_b|>L_0$ regime, it
is well described by the low energy CFT in the $d-1$ description.

As in \cite{{furtherinfl},{Larsen:2003pf}}, it would be interesting to go further and estimate the slow roll parameters in terms
of the rate of RG flow in the system.

\subsection{Reheating}

The most naive form of reheating in the bulk involves the inflaton
oscillating about the $\Lambda_0$ minimum and transferring its
energy to bulk fields before settling down at the minimum.  In the
$d-1$ description, this corresponds to a coupling oscillating in
time about a new fixed point value and injecting energy into the
CFT modes before settling down to the CFT fixed point value.  The
oscillation of the coupling changes the masses of the emerging
CFT$_0$ degrees of freedom, which results in particle production
generically. As in \cite{trapping}, as the coupling proceeds past
the CFT value, the CFT$_0$ degrees of freedom mass up again and
their energy density provides a restoring force trapping the
coupling near its fixed point value.  Hence inflation is dual to
moduli trapping.
This is especially precise if we make the choice of convention
discussed above where we take the same geometry everywhere in the
$3d$ spacetime, so that the entire effect is one of changing
species number (time dependent RG flow).

\subsection{Density perturbations}

Inflation produces structure in the universe via a beautifully simple application of quantum field theory in the
expanding universe.  In the original flat spatial coordinates (\ref{metric}), the mode solutions to the
equations of motion for the scalar perturbation grow in proper size in the $\vec x$ direction until they stretch
outside the horizon of the approximate de Sitter of Hubble constant $H_I$, at which point their amplitude
becomes constant. After the exit from inflation, the transition to $H_0$, these modes reenter the  horizon
and become dynamical again.  The two point Greens
function of these modes yields the Gaussian power spectrum seen in the CMBR observations.

It is interesting to ask how this process appears from our holographic screen.  The initial dynamics of modes
stretching toward the horizon takes forever from the point of view of the approximately static observer in the
initial inflationary phase, if we approximate this as a de Sitter phase.  In the radial direction, this
stretching toward the horizon corresponds to a thermalization in the dual CFT at low energies.  The next step,
the exit from inflation, entails a relatively sudden influx of new light species in the $d-1$ description.  This
means that an excitation which was nearly thermalized with respect to the $N_I$ species of the inflationary
phase, is suddenly far out of equilibrium with respect to the new light $N_0$ species that have appeared.  This
is the $d-1$ dimensional holographic description of the ``reentry" of modes into the horizon in the $d$
dimensional description.

\section{Conclusions}

In this paper we worked out some basic features of the UV regime of the dS/dS correspondence.  de Sitter space
is a particularly symmetric example of a Randall-Sundrum system with a similar holographic dual description at
the level of effective field theory.  Bulk gravity calculations allow one to probe some of the basic physics of
induced gravity above the scale where gravitational self-interactions become strong.

We computed several quantities determining how the lower dimensional theory behaves at energies above the
cutoff.  This resulted in a holographic verification that the total central charge and heat capacity is zeroed
out and that a simple asymptotic dressing of operator dimensions arises.  These are both features familiar in 2d
gravity plus matter systems.  Direct couplings between the two CFTs are also required.

The duality naturally extends to situations with changing cosmological constant, such as inflation and landscape
decays.  For these processes the $d-1$ dimensional holographic description involves a time dependent RG flow in
which new degrees of freedom become light in the transition.

Repeated application of our duality allows one to go to sufficiently low dimensions, that is two or one, so that
gravity becomes non-dynamical and its effects reduce to constraints. As long as the matter part shows no
pathologies, and the results of our current investigation suggest that it does not, one has a well defined
holographic dual in terms of a constrained quantum mechanical system. We briefly outlined how this then can be
used to construct observables for quantum gravity on backgrounds with accelerated expansion.

The motivation for this dual formulation is ultimately to provide a framework for the physics of accelerated
expansion in the real universe (both with respect to early inflation and late acceleration). Although at large
radius and low energies the effective weakly coupled description remains the bulk $d$ dimensional one, the
description in terms of $d-1$ dimensional physics may shed light on the physics of inflation and dark
energy.\footnote{For example the considerations in section 7 might provide a framework in which to address
questions about the relation of holography to CMBR measurements, questions explored for example in
\cite{hogan}.}

\section*{Acknowledgements:}

We would like to thank Ami Katz for collaboration during
intermediate stages of the project.  We would also like to thank
S. Giddings, J. Hartle, S. Kachru, N. Kaloper, E. Katz, A.
Maloney, and D. Tong for discussions.  The research of AK is
supported in part by DOE contract \#DE-FG03-96-ER40956 and AK is
also supported by NSF Grant SBE-0123552 as an ADVANCE professor.
E.S. is supported in part by the DOE under contract
DE-AC03-76SF00515 and by the NSF under contract 9870115.

\bibliography{gravds}
\bibliographystyle{utphys}
\end{document}